\documentclass[fleqn,usenatbib]{mnras}

\usepackage{newtxtext,newtxmath}

\usepackage[T1]{fontenc}

\DeclareRobustCommand{\VAN}[3]{#2}
\let\VANthebibliography\thebibliography
\def\thebibliography{\DeclareRobustCommand{\VAN}[3]{##3}\VANthebibliography}

\usepackage{graphicx}	
\usepackage{amsmath}	
\usepackage{longtable}

\title[Temporal variations of the rotation rates]{The rotation rate of solar active and ephemeral regions -- II. Temporal variations of the rotation rates}

\author[A. S. Kutsenko et al.]{
Alexander S. Kutsenko,\thanks{E-mail: alex.s.kutsenko@gmail.com (ASK)}
Valentina I. Abramenko
and Daria V. Litvishko
\\
Crimean Astrophysical Observatory, p/o Nauchny, Crimea, 298409, Russia
}

\date{Accepted XXX. Received YYY; in original form ZZZ}

\pubyear{2022}

\begin{document}
\label{firstpage}
\pagerange{\pageref{firstpage}--\pageref{lastpage}}
\maketitle

\begin{abstract}
Systematic studies of the rotation rate of sunspot groups using white-light images yield controversial results on the variations of the rotation rate: sunspot groups were found to either accelerate or decelerate systematically. This disagreement might be related to shortcomings of the method used to probe the rotation rate of sunspot groups. In contrast to previous works, in this study we use magnetic field maps to analyse the variations of the rotation rate of active regions. We found that an active region may exhibit either acceleration or deceleration during the emergence while the rotation rate remains almost unchanged during decay. Hence, we suppose that there is no systematic geometrical inclination to the radial direction of the apex of the subsurface magnetic flux loop forming an active region. A thorough comparison of the rotation rate of unipolar and bi/multipolar active regions revealed no significant changes in the rotation rate of decaying active regions. In contrast to previous works, we presume the rotation rate to keep constant (within the expected uncertainties) during the evolution of an active region after emergence.
\end{abstract}




\section{Introduction}
\label{sec_intro}

The differential rotation of the Sun was discovered almost 400 years ago by Christoph Scheiner who studied Galileo's sunspot observations \citep{Paterno2010}. A bench of works on the analysis of the solar rotation has been done since that time. The differential rotation is measured by tracking features on the solar surface as well as by analysing the Doppler velocities (the spectroscopic method) of solar plasma. The most sophisticated helioseismology approaches may shed light on the internal rotation of the convection zone \citep[e.g.][]{Howe2000}. We refer the reader to reviews by \citet{Beck2000} and \citet{Paterno2010} for more details.

The issue that is not well understood yet is the difference between the rotation rates derived using feature tracking and spectroscopic measurements. Thus, numerous studies show that, for instance, sunspot groups rotate systematically faster as compared to surface plasma at the same latitude. As an example, \citet{Howard1970} found that their spectroscopic measurements resulted in approximately 0.6 deg~d$^{-1}$ slower equatorial rotation rate as compared to tracer measurements of sunspots in \citet{Newton1951}. The same tendency can be seen in tables 1 and 2 in \citet{Beck2000} who made a comparison of different methods used to probe the differential rotation.

Since the datasets on sunspots are the most long-term ones, systematic studies of sunspot rotation was carried out by many authors. The results of these studies are controversial at some points. There exists a general agreement on the relationship between the rotation rate and size of a sunspot group: larger sunspot groups tend to rotate slower \citep[e.g.][]{Ward1966, Howard1984}. At the same time, there is no consensus on the dependence between the rotation rate and the age of a sunspot group. A solid number of researchers concluded that young sunspot groups rotate at a high rate while the rotation decelerates as the sunspot group ages \citep[e.g.][to mention a few]{Balthasar1982, Zappala1991, Pulkkinen1998, Ruzdjak2004}. On the other hand, \citet{Javaraiah1997, Hiremath2002} and \citet{Sivaraman2003} found sunspot groups, at least long-living ones, to accelerate their rotation with time.

\begin{figure}
	\includegraphics[width=\columnwidth]{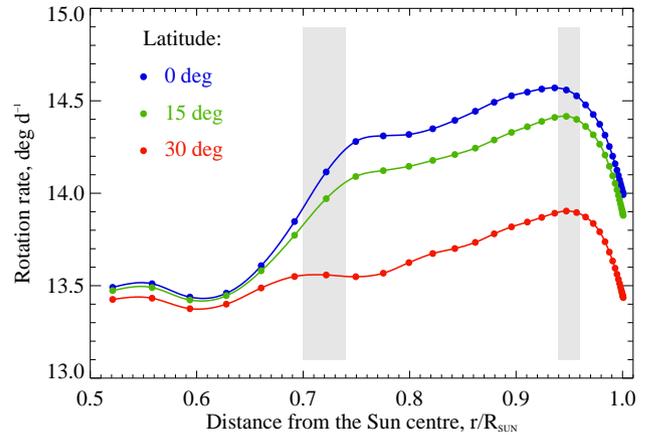}
	\caption{The rotation rate of plasma in the convection zone against the distance from the centre of the Sun. Grey rectangulars denote the approximate depths of the tachocline (0.72~R\sun) and of the leptocline (0.95 R\sun). The data are courtesy of Dr. Rachel Howe \citep{Howe2000}}.
	\label{fig_RHowe}
\end{figure}

The difference in the rotation rate of individual sunspot groups or, more generally, active regions is often explained in the framework of an anchoring hypothesis. This hypothesis assumes magnetic flux bundle to be anchored at some depth in the convection zone. As a consequence, the rotation rate at the surface is governed by the rotation velocity of plasma at the depth of anchoring. In such a case, the variations in the rotation rate of an active region can be interpreted as a rise of its magnetic ``roots'' through the convection zone. Note that, as we have already mentioned in the first paper of this series of works \citep{Kutsenko2021}, the anchoring hypothesis assumes the magnetic flux loop to emerge in the radial direction without significant perturbations. In Fig.~\ref{fig_RHowe} we present the rotation rates of plasma in the convection zone against the distance from the centre of the Sun \citep[the data are courtesy of Dr. Rachel Howe, see fig.~1 in][]{Howe2000}. One can see that for the 0--30 deg latitudinal zone the rotation rate increases from tachocline (0.72 R$_{\sun}$) to leptocline (0.95 R$_{\sun}$) and then decreases significantly near the photosphere. Assuming magnetic flux bundle ``roots'' to govern the rotation rate of an active region on the surface, acceleration or deceleration of the active region rotation rate may imply the rise of the magnetic ``roots'' from the tachocline or from the leptocline to the surface, respectively. These assumptions are often used to infer the initial depth of the formation of magnetic flux bundle that give birth to active regions \citep[e.g.][]{Hiremath2002, Brandenburg2005}.

It is worth noting that the anchoring hypothesis has no a solid theoretical foundation \citep[see section~3 in][]{Moradi2010}. Nevertheless, magnetic field lines have no ends and they must penetrate through the surface to the convection zone. Many global dynamo models implicitly assume that the magnetic field lines continue to and close near the base of the convection zone \citep[e.g.][]{Charbonneau2020}. Recent state-of-the-art numerical simulations of emerging magnetic flux bundle in \citet{Chen2017} support this idea. The authors found that magnetic flux bundles are still coherent structures down to tens of Mm below the photosphere, although their continuation mixes with other magnetic structures in the bulk of the convection zone. We will discuss this issue in the following sections.

The variations of the rotation rate of young sunspot groups may be also explained by different physical mechanisms not related to the anchoring \citep[e.g.][]{Petrovay2010}. For example, the geometry of the emerging flux bundle -- the inclination of the top part of emerging magnetic loop -- may result in artificial apparent acceleration or deceleration of a forming sunspot group. Proper motions of each individual footpoint may also yield changes in the rotation rate.

When interpreting the measured rotation rates of sunspot groups, one must keep in mind a very frustrating drawback in the methodology. This drawback was pointed out by \citet{Petrovay1993} who argued that more rapid decay of the following part of a sunspot group results in the artificial shift of the apparent area-weighted centre toward the leading part \citep[see fig.~2 in][]{Petrovay1993}. In \citet{Kutsenko2022} we evaluated the contribution of this effect to the measured rotation rate. For a set of 670 active regions we calculated the area-weighted centre positions in continuum intensity images (i.e. active regions were considered as sunspot groups) and geometrical centre positions using line-of-sight magnetograms. These centres were used to measure the rotation rate of active regions. The centre of a sunspot groups in continuum intensity images usually exhibited a shift toward the leading part of the group during the decay of the following part. As the following spot decayed completely, the shift ceased and the centre of the sunspot group coincided with the centre of the leading part \citep[see fig.~1 in][]{Kutsenko2022}. As a result, one could observe artificial acceleration of the centre during the decay of the following part. As the following part decays completely, the centre remains near the leading part yielding apparent deceleration of the sunspot group. Obviously, we can detect both false acceleration and false deceleration. In \citet{Kutsenko2022} we found that the rotation rate inferred from continuum intensity images is on average 0.45 deg~d$^{-1}$ higher than that inferred from magnetograms for the same set of active regions. We attribute this difference exclusively to faster disappearance of the following part in continuum intensity images. Recall that the difference between the rotation of the Sun near the equator and at 30 deg latitude, i.e. the zone where most active region emerges on the solar surface, is of about 0.5 deg~d$^{-1}$. This value is comparable to the average difference between the rotation rates derived from continuum intensity images and from magnetograms.

Our goal in this work is to analyse the variations of the rotation rates of active regions using magnetic field data. The discrepancies in the rotation rate variations found for old and young sunspot groups as described above might be related to the drawbacks of the data applied and of the measurement techniques. We insist that, in contrast to continuum intensity images, magnetograms ensure better evaluation of an active region geometry providing real positions of the leading and following parts of an active region. Consequently, in our opinion, magnetograms yield more robust measurements of the rotation rates.

In \citet{Kutsenko2021} we analysed averaged rotation rates of active and ephemeral regions using line-of-sight magnetograms provided by {\it Helioseismic and Magnetic Imager} \citep[HMI,][]{Schou2012} on board the {\it Solar Dynamics Observatory} \citep[SDO,][]{Pesnell2012}. We confirmed the inference made by many authors regarding the rotation of active regions: the weakest ephemeral regions exhibited the fastest rotation while large active regions with high magnetic flux rotated significantly slower \citep[see also, e.g.,][]{Lamb2017}. However, we also found unipolar active regions to disobey this rule. Unipolar active regions exhibited relatively low magnetic flux and rotated slowly at a rates of large mature active regions. In this work we propose our explanation of this issue.

In Section~\ref{sec_data} we describe the data reduction techniques and methods used to probe the rotation rates. Section~\ref{sec_emerging} presents our results on the rotation rate analysis of emerging active regions. In Section~\ref{sec_unipolar} we discuss the rotation of decaying unipolar and recurrent active regions. Finally, in Section~\ref{sec_concl} we summarize our results and propose a qualitative physical mechanisms that can explain our findings.

\section{Data and Method}
\label{sec_data}

In this work we used the data prepared in \citet{Kutsenko2021} and we refer the reader to that paper for details. Briefly, we used cubes of data extracted from 4096$\times$4096 pixels line-of-sight full-disc SDO/HMI magnetograms with the pixel size of 0.5$\times$0.5~ arcsec$^{2}$ and spatial resolution of 1~arcsec. The cubes of data represented temporal evolution of magnetic field of individual active regions. Each active region was initially manually bounded at full-disc magnetograms and then tracked back and forth in time in the consecutive magnetograms. The cadence of the data was 720~s. In all, we prepared the cubes of data for 864 active regions observed between 2010 and 2016.

The flux-weighted centres were calculated for each magnetic polarity of an active region separately. Before this procedure, the magnetograms were binned by 2$\times$2 pixels in order to diminish the influence of small-scale magnetic features. Only pixels with magnetic flux density exceeding 100~Mx~cm$^{-2}$ by modulus were used in the calculations. The centre of an active region was defined as the geometrical centre between the flux-weighted centres of the opposite magnetic polarities.

The active region centre in coordinates of magnetogram pixels was converted to heliographic Stonyhurst coordinates using World Coordinate System library in \textsc{IDL SolarSoft} package. The longitude versus time profiles were fitted by a linear approximation to measure the synodical rotation rate of an active region. The sidereal rotation rate, $\omega_{sid}$, was calculated by adding a term to the derived synodical rotation rate. The term accounts for the relative rotation of the Earth with respect to the Sun. The details of the calculation of this term can be found in \citet{Skokic2014, Lamb2017, Kutsenko2021}.

\section{The rotation rate of emerging active regions}
\label{sec_emerging}

We focused on the variation of the rotation rate of an active region from the very emergence. We selected 65 emerging active regions from our set using the following criteria: i) each active region must emerge within quiet-Sun area with no significant pre-existing magnetic flux; ii) there must be predominantly a single occurrence of emergence. The latter criterium is very important since additional emergence in a well-formed active region may shift the flux-weighted centre of one of magnetic polarity resulting in the shift of the centre of the entire active region. In general, we found that even emergence or appearance in the field-of-view of a small portion of a new magnetic flux may vary the position of the centre and, consequently, the rotation rate of the active region significantly. This effect is especially relevant for weak active regions. In addition, a rapid decay of a coherent magnetic flux concentration into a network field may also cause essential variations in the apparent rotation rate. Although it is not a trivial task to properly evaluate the uncertainties in the measured rotation rates, based on our experience, we suppose the uncertainties to be of order of 0.3 deg~d$^{-1}$ for large active regions. The list of the analysed active regions is presented in Table~\ref{table1} in the Appendix.

\begin{figure*}
	\includegraphics[width=\linewidth]{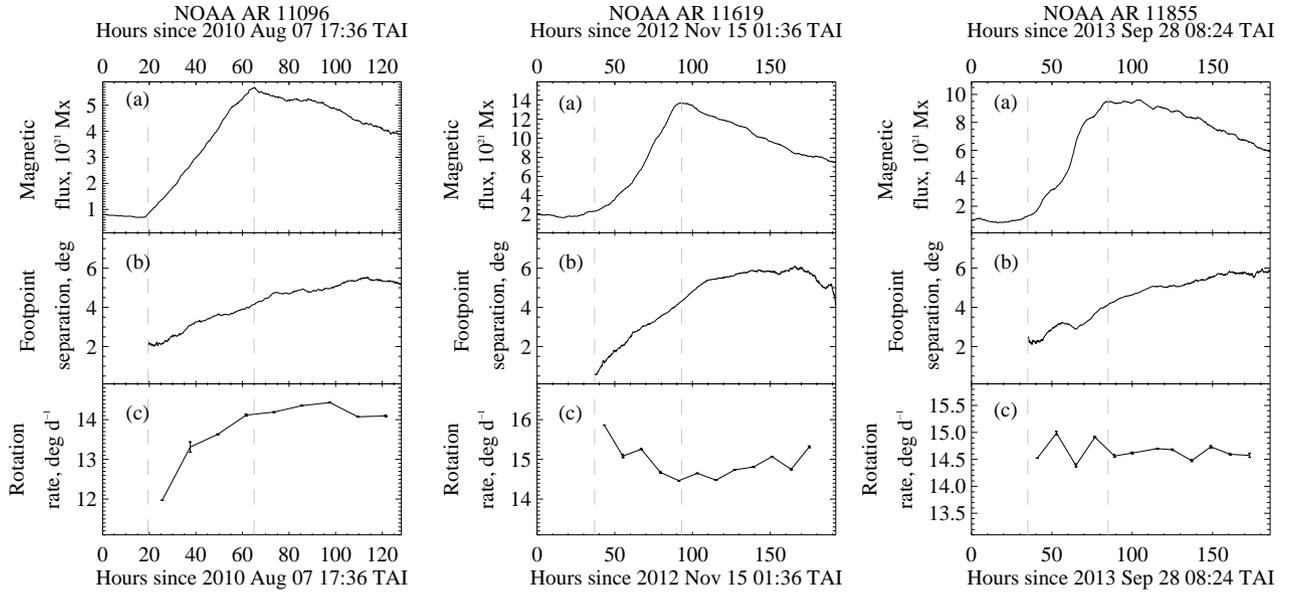}
	\caption{The variations of the total unsigned magnetic flux (panels a), footpoint separation (panels b), and rotation rate (panels c) of NOAA active regions 11096 (left), 11619 (middle), and 11855 (right). Bottom panels show typical patterns of the rotation rate in active region during emergence: acceleration (left), deceleration (middle), and slight fluctuations around some mean value. In most cases the rotation rate remains quasi-constant during active region decay. Vertical dashed lines point the times of the emergence onset and of the peak magnetic flux. Error bars in the bottom panels show the uncertainties of the linear approximations applied to derive the rotation rate.}
	\label{fig_profiles}
\end{figure*}

We found that a well-known 12/24-hour artificial periodicity in the SDO/HMI magnetic field data \citep[e.g.][]{Liu2012, Kutsenko2016} is readily pronounced in the measured rotation rate. That is why, in order to measure the ``instantaneous'' rotation rate, we fitted the longitude-versus-time profile by a linear approximation within a 12-hour wide rectangular window. The typical results are shown in Fig.~\ref{fig_profiles}. Upper panels (a) in the figure show the variations of the total unsigned magnetic flux of three active regions. Panels (b) show the footpoint separation (in heliographic degrees) that was calculated as the distance between the flux-weighted centres of opposite magnetic polarities. The measured rotation rates of active regions at a cadence of 12~h are shown in panels (c) of Fig.~\ref{fig_profiles}.

We found that the patterns of the rotation rate variation in the analysed active regions could be divided into three sets. First, the rotation rate increases (acceleration) during the active region emergence (15 cases out of 65). A typical example of this pattern is shown in the left column of Fig.~\ref{fig_profiles}. Second, an active region slows down (deceleration) as the emergence proceeds (the middle column of Fig.~\ref{fig_profiles}). Such a pattern was revealed in 25 active regions. Finally, the rotation rate varies around some mean value without any well-pronounced trend (25 cases, see Fig.~\ref{fig_profiles}, right column).  We will refer to these latter cases as stationary ones through the rest of the text. The revealed pattern of the rotation rate variation is indicated in the second column of Table~\ref{table1}. Interestingly, the number of active regions in each set is similar.

Note that in most cases the variations of the rotation rate after the emergence is insignificant with respect to the variations during the emergence stage. Beside, although the footpoint separation may keep increasing after the observed peak magnetic flux (Fig.~\ref{fig_profiles}c, left and right column), this expansion of the magnetic flux bundle does not affect significantly the rotation rate. Seemingly, after the emergence complete, further separation of magnetic footpoints occurs symmetrically with respect to the geometrical center of the magnetic loop.


\begin{figure*}
	\includegraphics{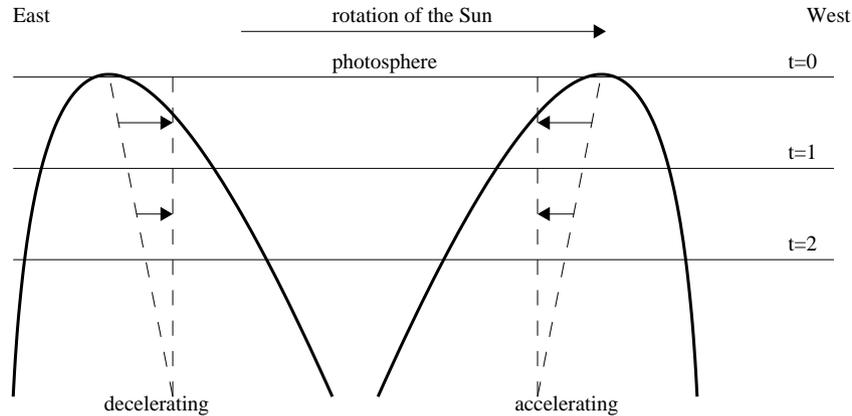}
	\caption{A simplified sketch of the possible inclination of the magnetic loop apex prior to emergence. An eastward (westward) inclination of the loop apex in the left (right) side of the figure results in a proper motion of the geometrical center of the loop with respect to the outbreak position in the West (East) direction. As the less inclined part of the loop breaks to the surface, the proper motion slows down resulting in apparent deceleration (acceleration) of the active region at the surface.}
	\label{fig_sketch}
\end{figure*}

An explicit pattern (acceleration/deceleration) in the rotation rate during emergence might imply eastward of westward geometrical inclination to the radial direction of the entire emerging magnetic flux loop. This issue can be understood in a simplified sketch shown in Fig.~\ref{fig_sketch}. The sketch shows two possible cases of emerging magnetic loop with inclined apex. In addition to the rotation of the loop as a whole, in the case of Eastward inclination (the left loop in Fig.~\ref{fig_sketch}), the geometrical center (bisector) of the two footpoints moves in the Westward direction from the outbreak position as the loop emerges \citep[cf. fig.~14 in ][]{Caligari1995}. As a consequence, the apparent rotation of the active region is faster as compared to ``real'' Westward rotation of the loop base. As the emergence proceeds, the symmetrical part of the loop probably rises to the surface. The proper motion of the footpoints halts and the active region starts to rotate slower at some constant rate. The observer will detect the deceleration of the rotation rate of the active region. It is trivial to see that in the case of the initial Westward inclination of the apex of the magnetic loop, one will observe the acceleration of the active region.

The geometrical East-West asymmetry of a magnetic loop emerging through the convection zone is readily obtained in simulations \citep[see, e.g., Section~5 in][]{Fan2021}. Thus, \citet{MorenoInsertis1994} and \citet{Caligari1995} performed a number of MHD simulations of the emergence of a thin flux tube through the convection zone in a spherical domain. The flux tube, initially stored in the overshoot region, forms an emerging loop as a result of undular instability. Due to the angular momentum conservation, the Coriolis force makes the summit of the rising loop move retrograde, i.e. in the direction opposite to the rotation of the Sun. As a result, a geometrical eastward inclination of the loop appears with the leading part of the loop being more horizontal \citep[see fig.~3 in][]{Caligari1995}. At the same time, during the emergence, plasma tends to flow along the tube from the apex to deeper layers. Again, due to the angular momentum conservation, the Coriolis force additionally drives the plasma from the leading part of the loop through the apex to the following part \citep{Fan1993}. More evacuated leading part becomes more buoyant and may rise faster to the surface. This effect may result in a more vertical inclination of the leading part as compared to the following leg \citep[see fig.~8 in][]{Fan2008}. Seemingly, the Eastward bending of the magnetic loop by the Coriolis force and Westward inclination of the loop due to fast rising of a more buoyant leading leg might act simultaneously. We may suppose the final inclination of the loop to be a result of a competition of these two processes with the non-inclined loop geometry as an option.

\citet{Caligari1995} argued that the analysis of the proper motions of the leading and following polarities may confirm the subsurface inclination of the magnetic loop. Thus, \citet{vanDrielGesztelyi1990} supposed emerging magnetic flux loops to be tilted eastward: the leading part of a sunspot usually exhibited faster proper motion with respect to Carrington longitude. More recent study by \citet{Schunker2016} also revealed faster proper motion of the leading magnetic polarity relative to Carrington rotation rate as compared to the following one. However, when considering the rotation with respect to the differential rotation of the surface plasma, \citet{Schunker2016} concluded that both magnetic polarities separate symmetrically in the east-west direction. In contrast to works mentioned above, we compared the rotation rate of an active region during emergence to that during decay.

The plasma flow from the leading to the following leg of an emerging magnetic loop, which was described above, causes asymmetry in the magnetic flux density in the following and leading polarities of an active region: the leading polarity is usually stronger and more coherent. The inclination of the loop leg might affect the magnetic flux density in the leg as well. From the general point of view, assuming constant magnetic flux within the tube, in the more horizontal magnetic leg the flux is distributed over the larger area resulting in the lower mean magnetic flux density. On the contrary, the more vertical leg presumably must exhibit higher magnetic flux density. If our speculations are correct, the eastward-inclined (decelerating) magnetic loops with more vertical following leg should exhibit lower mean flux density in the preceding part and higher mean flux density in the following part as compared to westward-inclined accelerating magnetic loops.

\begin{figure*}
	\includegraphics{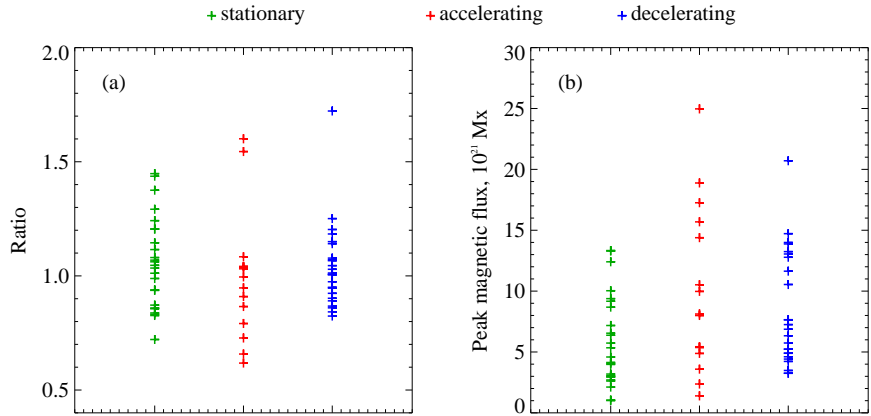}
	\caption{Left -- The distributions of the ratio between the mean magnetic flux density in the preceding polarity and that in the following polarity (see text) for stationary (green crosses), accelerating (red crosses), and decelerating (blue crosses) emerging active regions. Right -- The distributions of the peak magnetic flux for stationary (green crosses), accelerating (red crosses), and decelerating (blue crosses) emerging active regions.}
	\label{fig_densities}
\end{figure*}

For the active regions in our sample, we calculated the ratio between the mean magnetic flux density in the preceding polarity and that in the following polarity (the last column in Table~\ref{table1}). In the calculations we used pixels with absolute magnetic flux density exceeding 100 Mx~cm$^{-2}$. The ratio exceeding unity implies higher mean magnetic flux density in the preceding polarity of the loop. The results are presented in the left panel (a) of Fig.~\ref{fig_densities}. The pattern of the rotation rate variations is coded by colour. The visual analysis hints that the distribution of accelerating active regions is shifted toward a lower ratio as compared to that for decelerating ones. On one hand, this result is opposite to what is expected in our simplified considerations. On the other hand, the mean values of the ratio are 1.06$\pm$0.20 for stationary active regions, 0.99$\pm$0.28 for accelerating active regions, and 1.04$\pm$0.18 for decelerating active regions. Thus, from the statistical point of view, the distributions are the same. Besides, the underlying physics is definitely more complicated and our naive expectations might be far from reality.

\citet{Caligari1995} also argued that the total magnetic flux should affect the inclination of the emerging magnetic loop with stronger loops (in the sense of the total magnetic flux) being more inclined eastward. Then, eastward-inclined decelerating active regions should exhibit higher magnetic flux. The distributions of the peak magnetic flux in the sets of our active regions are shown in Fig.~\ref{fig_densities}b (see column~4 in Table~\ref{table1}). Indeed, stationary, presumably not inclined active regions exhibit the lowest average peak total magnetic flux (5.8$\pm$3.7)~10$^{21}$~Mx. Accelerating active regions exhibiting on average the highest peak total magnetic flux (10.0$\pm$6.9)~10$^{21}$~Mx as compared to (8.4$\pm$4.6)~10$^{21}$~Mx in decelerating active regions. Nevertheless, the distributions are statistically indistinguishable to draw a reliable conclusion regarding the relationship between the peak flux and inclination in a magnetic loop.

\section{The rotation rate of decaying active regions}
\label{sec_unipolar}

The analysis of the rotation rate of emerging active regions in Section~\ref{sec_emerging} did not allow us to make any conclusive decision regarding the rotation rate behaviour after the maximum magnetic flux. In most active regions the rotation rate remained almost constant with some negligible changes around some mean value (Fig.~\ref{fig_profiles}). This result might be caused by a specific selection of emerging active regions. Thus, we selected active regions emerging completely within 60 deg from the central meridian. Consequently, large active regions with peak unsigned magnetic flux exceeding 10$^{22}$~Mx were very scanty and did not get into our data set.

Perhaps, there exist some systematic variations of the rotation rates masked by relatively large uncertainties in the measurements for small active regions. In such a case these variations probably could be revealed in large active regions at timescales of several weeks. Unfortunately, our observations of a single active region are mostly limited to approximately 10-12 days when the active region could be observed at the visible solar disc. On the other hand, we can study many active regions at different stages of their life captured by SDO/HMI. As another option, we can explore recurrent active regions.

For the further analysis, we used data on rotation rates of 864 active regions studied in \citet{Kutsenko2021}. The rotation rate in that study was measured as an averaged value during the maximum development of an active region. That is, in contrast to bottom panels in Fig.~\ref{fig_profiles}, a single value of the rotation rate was derived for each tracer.

\begin{figure}
	\includegraphics[width=\columnwidth]{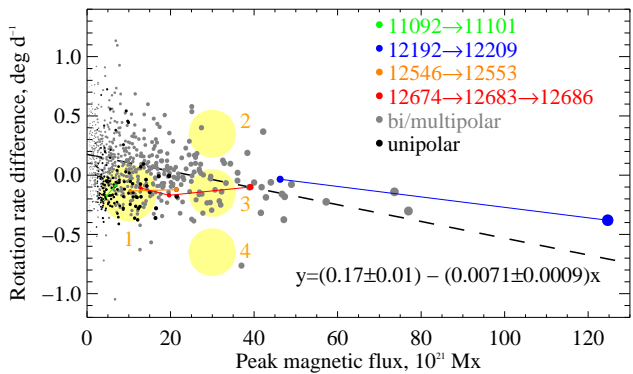}
	\caption{The difference between the rotation rate of individual active regions and the rotation rate derived using expression~(\ref{eq_difrot}). Bi- and multipolar active regions are shown in grey circles while unipolar active regions are shown in black circles. The area of the circle is proportional to the peak magnetic flux of an active region. Four recurrent active regions are shown in small coloured circles. Dashed line shows the best linear fitting of the distribution of bi/multipolar active regions. The slope of the line is shown by an expression in the plot. A yellow circle marked with \textit{1} shows the zone where the most of unipolar active regions are located. Yellow circles marked with \textit{2}-\textit{4} denotes possible zones unipolar active regions might evolve from.}
	\label{fig_difference}
\end{figure}

A direct comparison of the rotation rates of a large set of active regions is not appropriate due to strong latitudinal dependency of the rotation rate. As a possible solution one can consider only a set of tracers observed within a narrow latitudinal belt. However, in this case the number of points in each set might be too small to reveal some trends reliably. As another option, we can analyse the difference between the rotation rate of individual tracer at a certain latitude and some typical rotation rate at this latitude. As the typical rotation rate we adopted the empirical differential rotation law derived in \citet{Kutsenko2021}:
\begin{equation}
    \omega^{th} = 14.369 - 2.54 \sin^{2} \phi - 1.77 \sin^{4} \phi,
\label{eq_difrot}
\end{equation}
where $\phi$ is the heliographic latitude and the rotation rate $\omega^{th}$ is measured in deg~d$^{-1}$. The differences between the rotation rate and $\omega^{th}$ against the active region peak magnetic flux are plotted in Fig.~\ref{fig_difference}. A positive difference in Fig.~\ref{fig_difference} implies that the active region rotates faster than some average rotation rate at this particular latitude. The size of the data points in Fig.~\ref{fig_difference} is proportional to the peak magnetic flux of the active region. A dashed black line in the plot shows the best linear fittings of the distribution (unipolar active regions were excluded from the fitting). The slope of the fitting  as well as the visual analysis of the plot confirms clearly our previous deduction regarding slower rotation of larger active regions. 

In Fig.~\ref{fig_difference} we marked specifically the data for four recurrent active regions (color coding). Each small circle of a certain colour represents the same active region during consecutive solar rotations. The number of these tracers is not numerous in our data set. However, a clear tendency can be seen: while the total magnetic flux decreases with time, the rotation rate does not change significantly (the variation is of order of our expected uncertainty). Tracers labelled with small coloured circles shift to the left-hand side without considerable vertical displacement.

Unipolar active regions are shown in black circles in Fig.~\ref{fig_difference}. One can see that most of unipolar tracers are located within yellow circle marked with \textit{1}. The data points lie mostly below the linear approximation of the distribution (dashed black line). Although unipolar active region exhibit on average relatively low magnetic flux, their rotation rate is significantly slower than expected for the rest of active regions. We will discuss this issue in more detail in the next section.

\section{Conclusions and discussion}
\label{sec_concl}

In this work we analysed the temporal variations of the rotation rate in 65 emerging active regions using magnetic field data provided by SDO/HMI. We found no systematic pattern of the rotation rate variations in emerging active regions. Approximately one third of all tracers exhibited acceleration during the emergence phase (left panels in Fig.~\ref{fig_profiles}) while another third demonstrated rotation rate deceleration (middle panels in Fig.~\ref{fig_profiles}). Although the footpoint separation kept increasing, the deceleration or acceleration mostly halted as the active region gained peak magnetic flux. For the rest part of the tracers (right panels in Fig.~\ref{fig_profiles}) the rotation rate fluctuated around some mean value.

A systematic geometrical asymmetry in the East-West direction of the emerging magnetic loop apex might result in a certain systematic pattern of the rotation rate changes: acceleration or deceleration as the apex rises and stabilisation of the rotation rate as the magnetic loop straightens. However, we observe three possible patterns almost  equiprobably. The analysis of the mean magnetic flux density in the preceding and following polarities of active regions did not lead to a unique conclusion. We suppose that each individual magnetic flux loop forming an active regions may possess its own East-West subphotospheric inclination that determines the rotation rate pattern in the emerging stage. We do not exclude an option that the entire accelerating or decelerating patterns revealed in this work are caused by large uncertainties in the rotation rate calculations. Indeed, despite our strict criteria for emerging active region selection, several episodes of magnetic flux injection during the emergence of a single magnetic dipole are often observed. Beside, the pre-existing magnetic flux in the area of emergence is not perfectly zero. All these reasons may cause significant uncertainties in the determination of the active region centre and, as a consequence, of the rotation rate.

We also re-analysed the rotation rates of 864 active regions measured in \citet{Kutsenko2021}. We identified and visualised data for four recurrent active regions. Although the number of such tracers was too small, all of the analysed active regions exhibited almost unchanged rotation rates during several solar rotations (coloured circles in Fig.~\ref{fig_difference}). We presume that the rotation rate might remain stable during the entire life of an active region. Unfortunately, Solar Cycle 24 was not a strong cycle and we could not identify a solid number of recurrent active regions. Beside, recurrent active regions tend to emerge within active longitudes where the additional emergence of magnetic flux often occurs. The emergence of a new magnetic flux within pre-existing active region makes the measurements of rotation rate unreliable and we cannot use these tracers in our analysis.

In order to mitigate a low number of recurrent active regions in Solar Cycle 24, we explored 176 unipolar tracers in our set. We cannot observe an individual active region during its entire life due to solar rotation. On the other hand, we can observe a lot of active regions that are at a certain stage of their life. Unipolar active regions are definitely ``old'' active regions that lost at least a half of their peak magnetic flux. Hence, during their evolution, these tracers shifted to the left from their initial position on the difference-versus-flux plot shown in Fig.~\ref{fig_difference}. There are three options for the vertical displacement of the data points in the plot:
\begin{itemize}
    \item an active region decelerates with time and shifts downward in the plot in Fig.~\ref{fig_difference}. In the framework of the anchoring hypothesis, this pattern implies that the magnetic flux bundle that emerges as an active region was generated and rooted within the leptocline (Fig.~\ref{fig_RHowe}). As the active region evolves, its roots rise to the near-surface layers rotating at a lower rate. In such a case, most of active regions must be initially located within a yellow circle labelled with \textit{2}. For zones marked with yellow circles in Fig.~\ref{fig_difference}, we set the change of the rotation rate by 0.5~deg~d$^{-1}$ that corresponds to the difference between the plasma rotation rates within the leptocline and just below the solar surface (see Fig.~\ref{fig_RHowe});
    \item an active region exhibits almost constant rotation rate and shifts to the left without significant changes in the vertical direction in Fig.~\ref{fig_difference}. In this case, most tracers are expected to be initially concentrated within the yellow circle marked with \textit{3};
    \item finally, the rotation rate of a tracer may increase: the data point moves upward and to the left in Fig.~\ref{fig_difference} from the zone marked with yellow circle \textit{4} to the zone marked with yellow circle \textit{1}. Again, in the framework of the anchoring hypothesis, acceleration means generation and ``rooting'' of the magnetic flux bundle near the base of the convection zone (near the tachocline, Fig.~\ref{fig_RHowe}) and gradual rise of the roots toward the surface.
\end{itemize}

One can see in Fig.~\ref{fig_difference} that zone \textit{3} is the most populated one. Consequently, we suppose the \textit{3}$\longrightarrow$\textit{1} transition to be the most probable situation. Hence, the rotation rate presumably does not change during active region evolution after the peak magnetic flux. This assumption explains the slow rotation rate of unipolar active regions: these tracers rotate at a rate of mature large active regions that gradually evolved into unipolar magnetic structures.

We insist that magnetographic data allowed us to measure the rotation rate more precisely as compared to white-light observations. In our opinion, the variations in the rotation rate of sunspot groups found in previous studies (see Introduction \ref{sec_intro}) are due to measurement shortcomings \citep{Petrovay1993, Kutsenko2022}.

The unresolved problems that still exist are i) the physical connection between the rotation rate and peak magnetic flux of an active region, and ii) the difference between the rotation rate of non-magnetic plasma and magnetic structures at the surface. In accordance with the anchoring hypothesis, we presume the rotation rate of magnetic tracers to be governed by the ``roots'' located at some anchoring depth. In contrast to previous deductions, {we found the rotation rates to be unchanged with the active region evolution}, implying the constant anchoring depth. Our assumption is supported by simulations of active region emergence performed by \citet{Chen2017}. The authors found magnetic flux bundles forming an active region to be coherent formations deep below the surface. Assuming the active region rotation rate to be determined by the depth of the ``roots'', we presume that larger active regions are generated and anchored deeper in the bulk of the convection zone within slower internal plasma layers.

The unchanged rotation rate of active regions assumes the existence of the connection between the surface magnetic structure and relatively deep anchoring layers during the entire evolution of an active region after emergence. On the other hand, some kind of disconnection definitely takes place after the emergence is completed: the surface dynamics of an active region changes remarkably \citep[see a brief list of observations in support of this statement in the Introduction in][]{Schussler2005}. Thus, initially coherent stable magnetic polarities start fragmenting. The dispersed magnetic flux is then transported by local plasma flows to the network magnetic field. \citet{Fan1994} proposed the concept of a dynamical disconnection when the upper part of the magnetic tube becomes dynamically disconnected from the deeper magnetic structures due to significant decrease of the magnetic field strength in the tube at some intermediate depth. This concept was further elaborated by \citet{Schussler2005} who argued that the dynamical disconnection occurs in several days after the emergence complete. In our case, slowly rotating unipolar or recurrent active regions exhibit stable rotation rate and coherent structure for weeks or months implying, again, the existence of some connection with deeper layers. In our opinion, in these long-living magnetic structures the dynamical disconnection occurs probably in several weeks after the emergence is completed. Indeed, according to \citet{Fan1994} and \citet{Schussler2005}, the most pronounced manifestation of the dynamical disconnection is the fragmentation of coherent magnetic structures at the surface. We analyse mostly coherent magnetic structures before they start to fragment into a network field. Therefore, the active region is still in the stage before disconnection.

\section*{Acknowledgements}

We deeply appreciate the help of the anonymous referee whose highly-professional comments motivated us to get a deeper insight in the subsurface dynamics of magnetic fluxes. We are cordially grateful to Dr. Rachel Howe for providing the data on the internal rotation rate of the Sun. Magnetic field data are courtesy of NASA/SDO and HMI science team. The study presented in Section~\ref{sec_emerging} was supported by the Russian Science Foundation, Project 18-12-00131.

\section*{Data Availability}

The SDO/HMI data that support the findings of this study are available in the JSOC (http://jsoc.stanford.edu/) and can be accessed under open for all data policy. Derived data products are available in the article \citet{Kutsenko2021} and from the corresponding author (ASK) on request.




\begin{thebibliography}{99}





\bibitem[\protect\citeauthoryear{Balthasar, Schuessler, \& Woehl}{1982}]
{Balthasar1982}
Balthasar H., Schuessler M., Woehl H., 1982, \solphys, 76, 21. doi:10.1007/BF00214127


\bibitem[\protect\citeauthoryear{Beck}{2000}]
{Beck2000}
Beck J.~G., 2000, \solphys, 191, 47. doi:10.1023/A:1005226402796


\bibitem[\protect\citeauthoryear{Brandenburg}{2005}]
{Brandenburg2005}
Brandenburg A., 2005, \apj, 625, 539. doi:10.1086/429584


\bibitem[\protect\citeauthoryear{Caligari, Moreno-Insertis, \& Schussler}{1995}]
{Caligari1995}
Caligari P., Moreno-Insertis F., Schussler M., 1995, \apj, 441, 886. doi:10.1086/175410


\bibitem[\protect\citeauthoryear{Charbonneau}{2020}]
{Charbonneau2020}
Charbonneau P., 2020, Liv. Rev. Sol. Phys., 17, 4. doi:10.1007/s41116-020-00025-6


\bibitem[\protect\citeauthoryear{Chen, Rempel, \& Fan}{2017}]
{Chen2017}
Chen F., Rempel M., Fan Y., 2017, \apj, 846, 149. doi:10.3847/1538-4357/aa85a0


\bibitem[\protect\citeauthoryear{Fan}{2008}]
{Fan2008}
Fan Y., 2008, \apj, 676, 680. doi:10.1086/527317


\bibitem[\protect\citeauthoryear{Fan}{2021}]
{Fan2021}
Fan Y., 2021, Liv. Rev. Sol. Phys., 18, 5. doi:10.1007/s41116-021-00031-2


\bibitem[\protect\citeauthoryear{Fan, Fisher, \& Deluca}{1993}]
{Fan1993}
Fan Y., Fisher G.~H., Deluca E.~E., 1993, \apj, 405, 390. doi:10.1086/172370


\bibitem[\protect\citeauthoryear{Fan, Fisher, \& McClymont}{1994}]
{Fan1994}
Fan Y., Fisher G.~H., McClymont A.~N., 1994, \apj, 436, 907. doi:10.1086/174967


\bibitem[\protect\citeauthoryear{Hiremath}{2002}]
{Hiremath2002}
Hiremath K.~M., 2002, \aap, 386, 674. doi:10.1051/0004-6361:20020276


\bibitem[\protect\citeauthoryear{Howard, Gilman, \& Gilman}{1984}]
{Howard1984}
Howard R., Gilman P.~I., Gilman P.~A., 1984, \apj, 283, 373. doi:10.1086/162315


\bibitem[\protect\citeauthoryear{Howard \& Harvey}{1970}]
{Howard1970}
Howard R., Harvey J., 1970, \solphys, 12, 23. doi:10.1007/BF02276562


\bibitem[\protect\citeauthoryear{Howe et al.}{2000}]
{Howe2000}
Howe R., Christensen-Dalsgaard J., Hill F., Komm R.~W., Larsen R.~M., Schou J., Thompson M.~J., et al., 2000, \sci, 287, 2456. doi:10.1126/science.287.5462.2456


\bibitem[\protect\citeauthoryear{Javaraiah \& Gokhale}{1997}]
{Javaraiah1997}
Javaraiah J., Gokhale M.~H., 1997, \aap, 327, 795


\bibitem[\protect\citeauthoryear{Kutsenko}{2021}]
{Kutsenko2021}
Kutsenko A.~S., 2021, \mnras, 500, 5159. doi:10.1093/mnras/staa3616


\bibitem[\protect\citeauthoryear{Kutsenko \& Abramenko}{2016}]
{Kutsenko2016}
Kutsenko A.~S., Abramenko V.~I., 2016, \solphys, 291, 1613. doi:10.1007/s11207-016-0940-z


\bibitem[\protect\citeauthoryear{Kutsenko \& Abramenko}{2022}]
{Kutsenko2022}
Kutsenko A.~S., Abramenko V.~I., 2022, OAst, 30, 219. doi:10.1515/astro-2021-0029


\bibitem[\protect\citeauthoryear{Lamb}{2017}]
{Lamb2017}
Lamb D.~A., 2017, \apj, 836, 10. doi:10.3847/1538-4357/836/1/10


\bibitem[\protect\citeauthoryear{Liu et al.}{2012}]
{Liu2012}
Liu Y., Hoeksema J.~T., Scherrer P.~H., Schou J., Couvidat S., Bush R.~I., Duvall T.~L., et al., 2012, \solphys, 279, 295. doi:10.1007/s11207-012-9976-x


\bibitem[\protect\citeauthoryear{Moradi et al.}{2010}]
{Moradi2010}
Moradi H., Baldner C., Birch A.~C., Braun D.~C., Cameron R.~H., Duvall T.~L., Gizon L., et al., 2010, \solphys, 267, 1. doi:10.1007/s11207-010-9630-4


\bibitem[\protect\citeauthoryear{Moreno-Insertis, Caligari, \& Schuessler}{1994}]
{MorenoInsertis1994}
Moreno-Insertis F., Caligari P., Schuessler M., 1994, \solphys, 153, 449. doi:10.1007/BF00712518


\bibitem[\protect\citeauthoryear{Newton \& Nunn}{1951}]
{Newton1951}
Newton H.~W., Nunn M.~L., 1951, \mnras, 111, 413. doi:10.1093/mnras/111.4.413


\bibitem[\protect\citeauthoryear{Patern{\`o}}{2010}]
{Paterno2010}
Patern{\`o} L., 2010, \apss, 328, 269. doi:10.1007/s10509-009-0218-0


\bibitem[\protect\citeauthoryear{Pesnell, Thompson, \& Chamberlin}{2012}]
{Pesnell2012}
Pesnell W.~D., Thompson B.~J., Chamberlin P.~C., 2012, \solphys, 275, 3. doi:10.1007/s11207-011-9841-3


\bibitem[\protect\citeauthoryear{Petrovay}{1993}]
{Petrovay1993}
Petrovay K., 1993, in Zirin H., Ai G.,Wang H., eds., ASP Conf. Ser. Vol. 46,
The Magnetic and Velocity Fields of Solar Active Regions. Astron. Soc.
Pac., San Francisco, p. 123


\bibitem[\protect\citeauthoryear{Petrovay \& Christensen}{2010}]
{Petrovay2010}
Petrovay K., Christensen U.~R., 2010, \ssr, 155, 371. doi:10.1007/s11214-010-9657-8


\bibitem[\protect\citeauthoryear{Pulkkinen \& Tuominen}{1998}]
{Pulkkinen1998}
Pulkkinen P., Tuominen I., 1998, \aap, 332, 748


\bibitem[\protect\citeauthoryear{Ru{\v{z}}djak et al.}{2004}]
{Ruzdjak2004}
Ru{\v{z}}djak D., Ru{\v{z}}djak V., Braj{\v{s}}a R., W{\"o}hl H., 2004, \solphys, 221, 225. doi:10.1023/B:SOLA.0000035066.96031.4f


\bibitem[\protect\citeauthoryear{Schou et al.}{2012}]
{Schou2012}
Schou J., Scherrer P.~H., Bush R.~I., Wachter R., Couvidat S., Rabello-Soares M.~C., Bogart R.~S., et al., 2012, \solphys, 275, 229. doi:10.1007/s11207-011-9842-2


\bibitem[\protect\citeauthoryear{Schunker et al.}{2016}]
{Schunker2016}
Schunker H., Braun D.~C., Birch A.~C., Burston R.~B., Gizon L., 2016, \aap, 595, A107. doi:10.1051/0004-6361/201628388


\bibitem[\protect\citeauthoryear{Schunker et al.}{2019}]
{Schunker2019}
Schunker H., Birch A.~C., Cameron R.~H., Braun D.~C., Gizon L., Burston R.~B., 2019, \aap, 625, A53. doi:10.1051/0004-6361/201834627


\bibitem[\protect\citeauthoryear{Sch{\"u}ssler \& Rempel}{2005}]
{Schussler2005}
Sch{\"u}ssler M., Rempel M., 2005, \aap, 441, 337. doi:10.1051/0004-6361:20052962



\bibitem[\protect\citeauthoryear{Sivaraman et al.}{2003}]
{Sivaraman2003}
Sivaraman K.~R., Sivaraman H., Gupta S.~S., Howard R.~F., 2003, \solphys, 214, 65. doi:10.1023/A:1024075100667


\bibitem[\protect\citeauthoryear{Skoki{\'c} et al.}{2014}]
{Skokic2014}
Skoki{\'c} I., Braj{\v{s}}a R., Ro{\v{s}}a D., Hr{\v{z}}ina D., W{\"o}hl H., 2014, \solphys, 289, 1471. doi:10.1007/s11207-013-0426-1


\bibitem[\protect\citeauthoryear{van Driel-Gesztelyi \& Petrovay}{1990}]
{vanDrielGesztelyi1990}
van Driel-Gesztelyi L., Petrovay K., 1990, \solphys, 126, 285. doi:10.1007/BF00153051


\bibitem[\protect\citeauthoryear{Ward}{1966}]
{Ward1966}
Ward F., 1966, \apj, 145, 416. doi:10.1086/148783

\bibitem[\protect\citeauthoryear{Zappala \& Zuccarello}{1991}]
{Zappala1991}
Zappala R.~A., Zuccarello F., 1991, \aap, 242, 480

\end{thebibliography}




\appendix

\section{Emerging active regions under study}
\label{appendix}
\onecolumn
\begin{center}
	\begin{longtable}{cccrc}
	\caption{The list of emerging active regions analysed in Section~\ref{sec_emerging}. The NOAA numbers of the active regions are listed in column 1. The letter in column 2 represents the type of rotation rate variation of an active region during the emergence: ``s'' stands for stationary rotation rate, ``a'' stands for acceleration, and ``d'' stands for deceleration. The dates of the observed peak magnetic flux in YYYY.MM.DD HH:MM format are listed in column 3, while the values of the measured peak magnetic fluxes are listed in column 4. Column 5 shows the ratio between the mean magnetic flux density in the preceding polarity and that in the following polarity (see Section~\ref{sec_emerging}).}
	\label{table1}

\\ \hline
NOAA & Type & Date & Peak magnetic& Ratio \\
	 &      &(TAI) & flux, 10$^{21}$ Mx & \\ \hline
\endfirsthead
11066& a& 2010.05.03 15:24&     2.4&     1.042\\
11096& a& 2010.08.10 16:12&     5.4&     0.791\\
11137& s& 2010.12.26 10:48&     2.1&     0.937\\
11143& s& 2011.01.07 19:00&     3.2&     1.115\\
11158& a& 2011.02.15 11:35&    25.0&     0.947\\
11173& s& 2011.03.17 17:24&     6.4&     1.080\\
11223& d& 2011.05.27 08:36&     5.2&     0.860\\
11241& d& 2011.06.25 22:36&     4.6&     0.842\\
11243& s& 2011.07.02 17:48&    13.3&     0.989\\
11247& d& 2011.07.10 19:00&     7.3&     0.924\\
11260& d& 2011.07.30 02:36&    20.7&     1.150\\
11266& a& 2011.08.09 02:12&     8.1&     0.909\\
ephemeral& s& 2011.08.26 17:36&     1.1&     1.145\\
11310& d& 2011.10.05 02:24&     3.3&     0.974\\
11311& a& 2011.10.05 01:12&     5.4&     1.601\\
11321& s& 2011.10.18 15:48&     4.0&     0.860\\
11334& d& 2011.10.31 21:36&     6.9&     1.013\\
11365& s& 2011.12.04 08:36&     8.7&     0.838\\
11372& a& 2011.12.10 11:00&     4.9&     1.038\\
11397& a& 2012.01.14 15:48&     3.6&     1.040\\
11400& a& 2012.01.15 03:36&     1.4&     0.618\\
11413& d& 2012.02.02 11:00&    10.5&     1.141\\
11416& a& 2012.02.12 05:00&    18.9&     0.866\\
11422& d& 2012.02.21 09:00&    13.2&     1.251\\
11444& s& 2012.03.25 04:24&     6.5&     1.448\\
11464& s& 2012.04.19 21:12&     1.0&     1.375\\
11465& a& 2012.04.23 09:00&    14.4&     1.031\\
11472& s& 2012.04.30 22:36&     5.7&     1.062\\
11490& d& 2012.05.30 04:00&    13.9&     1.005\\
11500& s& 2012.06.04 03:00&     2.7&     0.826\\
11512& s& 2012.06.28 04:36&    12.4&     1.292\\
11517& a& 2012.07.02 06:36&    10.5&     1.084\\
11533& s& 2012.07.29 03:00&     4.1&     1.437\\
11544& d& 2012.08.09 19:36&     6.3&     1.029\\
11551& d& 2012.08.21 14:00&     3.3&     0.950\\
11560& d& 2012.09.02 19:00&    14.7&     1.006\\
11561& s& 2012.08.31 11:48&     2.6&     0.856\\
11565& d& 2012.09.04 04:00&     5.2&     1.071\\
11568& s& 2012.09.09 21:00&     4.0&     0.832\\
11605& s& 2012.11.04 23:24&     3.0&     1.069\\
11619& d& 2012.11.19 05:48&    13.1&     1.203\\
11675& s& 2013.02.18 00:00&     5.3&     1.205\\
11723& d& 2013.04.17 01:48&    14.0&     1.067\\
11743& s& 2013.05.13 19:12&     9.2&     0.872\\
11747& d& 2013.05.17 03:36&     7.7&     1.008\\
11813& d& 2013.08.08 23:47&     4.4&     0.946\\
11817& s& 2013.08.14 14:36&    13.3&     1.242\\
11855& s& 2013.10.02 09:12&     9.4&     1.046\\
11887& d& 2013.11.05 14:12&    11.6&     1.045\\
11889& a& 2013.11.06 00:35&     8.0&     0.728\\
11911& s& 2013.12.02 00:23&     3.1&     0.721\\
11946& a& 2014.01.08 14:36&    15.7&     0.658\\
12018& d& 2014.03.27 22:00&     5.7&     0.824\\
12091& s& 2014.06.12 17:12&     2.9&     0.939\\
12098& d& 2014.06.25 22:36&     4.2&     0.868\\
12203& a& 2014.11.03 16:12&    10.0&     0.995\\
12234& d& 2014.12.12 23:24&     7.6&     1.722\\
12254& d& 2015.01.02 23:00&    12.8&     0.902\\
12266& s& 2015.01.20 05:48&    10.0&     1.012\\
12273& d& 2015.01.27 18:36&     4.9&     1.184\\
12336& d& 2015.05.06 03:00&     4.9&     1.078\\
12489& a& 2016.01.28 12:24&    17.3&     1.545\\
12579& s& 2016.08.23 17:12&     7.2&     1.206\\
12604& d& 2016.10.29 02:48&     3.5&     0.891\\
12614& s& 2016.11.29 20:24&     4.6&     1.034\\
	\end{longtable}
\end{center}



\bsp	
\label{lastpage}
\end{document}